\documentstyle[aps,amssymb]{revtex}
\begin{document}

\title{Creation of scalar and Dirac particles in the presence of a time varying electric
field in an anisotropic Bianchi I universe}
\author{V\'{\i}ctor M. Villalba\thanks{On leave from Centro de F\'{\i}sica, 
Instituto Venezolano de Investigaciones Cient\'{\i}ficas, IVIC 
Apdo 21827, Caracas 1020-A, Venezuela.
e-mail: villalba@th.physik.uni-frankfurt.de}\footnote{Alexander von Humboldt Fellow} 
and Walter Greiner} 

\address{Institut f\"ur Theoretische Physik, Universit\"at Frankfurt, \\
D-60054 Frankfurt am Main, Germany}

\maketitle

\begin{abstract}
In this article we compute the density of scalar and Dirac particles created by a cosmological
anisotropic Bianchi type I
universe in the presence of a time varying electric field. We show that the
particle distribution becomes thermal when one neglects the electric interaction.
\end{abstract}


\section{Introduction}

During the last decades a great deal of effort has been made in understanding
quantum processes in strong fields. Quantum field theory in the presence of
strong fields is in general a theory associated with unstable vacua. The
vacuum instability leads to many interesting features; among them particle
creation is perhaps the most interesting nonperturbative phenomenon.

In order to analyze the mechanism of particle creation in cosmological
backgrounds we have at our disposal different techniques such as the
adiabatic approach \cite{Parker,Fulling}, the Feynman path integral method 
\cite{Chitre}, the Hamiltonian diagonalization technique \cite
{Grib,Bukhbinder}, as well as the Green function approach \cite{Gitman}

The study of quantum processes in the presence of electromagnetic fields in
curved backgrounds has almost been restricted to constant electric fields. \
The density of particles created by an intense electric field was first
calculated by Schwinger \cite{schwinger} and different authors have discussed
this phenomenon in curved spaces. The technical difficulties associated with
solving the Klein-Gordon or Dirac equations in time dependent
electromagnetic fields have reduced the number of configurations studied  to
a few simple solvable models \cite{Grib}. A further difficulty arises when
one deals with cosmological backgrounds with singularities \cite{Fischetti,Sahni}. 
Here one cannot
apply the standard adiabatic approach and other equivalent methods developed
in the literature. It is worth mentioning that the presence of primordial
electric fields could enhance the particle creation mechanism and also
produce deviations from the thermal spectrum.

In order to discuss particle creation in an anisotropic universe, we
consider the cosmological scenario associated with the metric 
\begin{equation}
ds^{2}=-dt^{2}+t^{2}(dx^{2}+dy^{2})+dz^{2}  \label{1}
\end{equation}
The line element (\ref{1}) presents a singularity at $t=0$, The scalar
curvature is $R=2/t^{2}$, and consequently, the adiabatic approach \cite
{Parker} cannot be applied in order to identify particle states. Using the
Hamiltonian diagonalization method, Bukhbinder and coworkers 
\cite{Bukhbinder,bukh2,bukh3} 
were able to
compute the density of particles created by the background associated with
the metric (\ref{1}), and also in Ref. \cite{Odintsov1} this result was 
extended to include a time dependent electric field.

The purpose of the present article is to discuss the production of scalar
and Dirac particles in the anisotropic cosmological background associated with
the line element (\ref{1}) in the presence of a time dependent homogeneous
electric field. In order to compute the rate of particle creation we apply a
quasiclassical approach that has been used successfully in different scenarios\cite
{Lotze1,Lotze2,Villalba1,Villalba2}. The idea behind the method is the
following. First, we solve the relativistic Hamilton-Jacobi equation and,
looking at its solutions, we identify positive and negative frequency modes.
Second, we solve the Klein-Gordon and Dirac equations and, after comparing
with the results obtained for the quasiclassical limit, we identify the
positive and negative frequency states. The paper is structured as follows.
In Sec. II we solve the relativistic Hamilton-Jacobi equation and compute
the quasiclassical energy modes. In Sec. III we solve the Klein-Gordon
and Dirac equations and obtain the density of scalar and Dirac particles
created. The discussion of the results and final remarks are presented
in Sec. IV.

\section{Computation of the quasiclassical solutions}

The relativistic Hamilton-Jacobi equation can be written as
\begin{equation}
g^{\alpha\beta}\left(\frac{\partial S}{\partial x^{\alpha}}-eA_{\alpha
}\right) \left( \frac{\partial S}{\partial x^{\beta}}-eA_{\beta}\right)
+m^{2}=0
\end{equation}
where here and elsewhere we adopt the convention $c=1$ and $\hbar=1.$

The vector potential $A^{\mu }$ 
\begin{equation}
A^{\mu }=\frac{C}{t}\delta _{3}^{\mu }  \label{A}
\end{equation}
corresponds to a time dependent homogeneous electric field $C\hat{k}/t^{2}.$
The invariants 
\begin{equation}
\frac{1}{2}F^{\mu \nu }F_{\mu \nu }=-\frac{C^{2}}{t^{4}},\ F^{\mu \nu
}F_{\mu \nu }^{\ast }=0
\end{equation}
indicate that there is no magnetic field. Since the line element (\ref{1})
and the vector potential (\ref{A}) do not depend on the space variables we
can write the function $S$ as follows: 
\begin{equation}
S=\vec{k}\cdot \vec{r}+F(t).  \label{S}
\end{equation}
Substituting Eq. (\ref{S}) into the Hamilton-Jacobi equation one obtains 
\begin{equation}
\frac{k_{x}^{2}+k_{y}^{2}}{t^{2}}+(k_{z}-\frac{eC}{t})^{2}-(\frac{dF}{dt}%
)^{2}+m^{2}=0.  \label{dHJ}
\end{equation}
The solution of (\ref{dHJ}) has the following asymptotic behavior 
\begin{equation}
\lim_{t\rightarrow 0}F(t)=\pm \sqrt{k_{x}^{2}+k_{y}^{2}+e^{2}C^{2}}\log t
\label{pri}
\end{equation}
\begin{equation}
\Phi =e^{iS}\rightarrow C_{0}t^{\pm i\sqrt{k_{x}^{2}+k_{y}^{2}+e^{2}C^{2}}}
\label{zero}
\end{equation}
as $t\rightarrow 0,$ and 
\begin{equation}
\lim_{t\rightarrow +\infty }F(t)=\pm \sqrt{k_{z}^{2}+m^{2}}t\mp \frac{k_{z}eC%
}{\sqrt{k_{z}^{2}+M^{2}}}\log t,  \label{ter}
\end{equation}
\begin{equation}
\Phi =e^{iS}\rightarrow C_{1}e^{\pm i\sqrt{k_{z}^{2}+m^{2}}t}t^{\mp i\frac{%
k_{z}eC}{\sqrt{k_{z}^{2}+M^{2}}}}  \label{infi}
\end{equation}
as $t\rightarrow \infty .$ The positive or negative frequency mode is
selected depending on the sign of the operator $i\partial _{t}$. Positive
and negative frequency modes will have, respectively, positive and negative
eigenvalues; thus in Eqs. (\ref{pri})-(\ref{infi}) the upper signs are
associated with negative frequency modes and the lower signs correspond to
positive frequency states. After making this identification we proceed to
analyze the solutions of the Klein-Gordon and Dirac equations in the
cosmological background (\ref{1}).

\section{Particle production}       

\subsection{Scalar particles}

The covariant Klein-Gordon equation has the form 
\begin{equation}
g^{\alpha \beta }\left( \nabla _{\alpha }-ieA_{\alpha }\right) \left( \nabla
_{\beta }-ieA_{\beta }\right) \Phi -(m^{2}+\xi R)\Phi =0  \label{KG}
\end{equation}
where \ $\nabla _{\alpha }$ is the covariant derivative, $R$ is the scalar
curvature and $\xi $ is a dimensionless coupling constant which takes the
value $\xi =0$ in the minimal coupling case and $\xi =1/6$ when a conformal
coupling is considered. The scalar wave function $\Phi $ is normalized
according to the inner product \cite{birrel}
\begin{equation}
\left\langle \Phi _{1},\Phi _{2}\right\rangle =\int_{\sigma }(\Phi
_{2}\partial _{\mu }\Phi _{1}^{\ast }-\Phi _{1}^{\ast }\partial _{\mu }\Phi
_{2})\sqrt{-g}ds^{\mu }  \label{normkg}
\end{equation}
where $\sigma $ is an arbitrary spacelike hypersurface. Since the metric (%
\ref{1}) is stationary we choose a hypersurface orthogonal to the timelike
vector $t^{\mu }=\delta _{0}^{\mu }.$ Since Eq. (\ref{KG}) commutes with
each of the components of the linear momentum $\vec{p}=-i\nabla $, \ the
wave function $\Phi $ can be written as 
\begin{equation}
\Phi =t^{-1}\Delta (t)e^{i(k_{x}x+k_{y}y+k_{z}z)}.  \label{subs}
\end{equation}
Substituting Eq. (\ref{subs}) into Eq. (\ref{KG}) we reduce the problem of solving
\ Eq. (\ref{KG}) to that of finding a solution of the following second order
differential equation: 
\begin{equation}
\left( \frac{d^{2}}{dt^{2}}+\frac{k_{x}^{2}+k_{y}^{2}+e^{2}C^{2}+2\xi }{t^{2}%
}+k_{z}^{2}-\frac{2k_{z}eC}{t}+m^{2}\right) \Delta (t)=0  \label{urav}
\end{equation}
The solution can be expressed in terms of Whittaker functions \cite
{Lebedev,Abramowitz} 
\begin{equation}
\Delta ={\frak{C}_{1}}M_{k,\bar{\mu}}(\rho )+{\frak{C}_{2}} W_{k,\bar{\mu}}(\rho )
\label{sol}
\end{equation}
where $\frak{C}_{1}$ and $\frak{C}_{2}$ are arbitrary constants, and $\rho ,$ $k$
and $\bar{\mu}$ are given by the expressions 
\begin{equation}
\rho =-2i\sqrt{k_{z}^{2}+m^{2}}t,\ \bar{\mu}=i\sqrt{-\frac{1}{4}%
+k_{x}^{2}+k_{y}^{2}+e^{2}C^{2}+2\xi },k=-\frac{ik_{z}eC}{\sqrt{%
k_{z}^{2}+m^{2}}}
\end{equation}
Looking at the asymptotic behavior of $M_{k,\mu }(z)$ as $z\rightarrow 0$ 
\begin{equation}
M_{k,\mu }(z)\backsim e^{-z/2}z^{1/2+\mu },  \label{M}
\end{equation}
we find that, according to (\ref{zero}), and using the fact that all the
coefficients in Eq. (\ref{urav}) are real, the positive and negative
frequency solutions (\ref{sol}) at $t=0$ are 
\begin{equation}
\Delta _{0}^{+}={\frak{C}_{0}}^{+} M_{k,\bar{\mu}}(\rho ),\ \Delta _{0}^{-}=(%
{\frak{C}_{0}}^{+} M_{k,\bar{\mu}}(\rho ))^{\ast }={\frak{C}_{0}}^{+}(-1)^{-\bar{%
\mu}+1/2} M_{k,-\bar{\mu}}(\rho )
\end{equation}
where $\frak{C}_{0}^{+}$ is a normalization constant. Analogously, \ looking
at the behavior of $W_{k,\mu }(z)$ as $\left| z\right| \rightarrow \infty $
\begin{equation}
W_{k,\mu }(z)\backsim e^{-z/2}z^{k},  \label{W}
\end{equation}
we see that the corresponding positive and negative frequency modes as $%
\left| \rho \right| \rightarrow \infty $ are 
\begin{equation}
\Delta _{\infty }^{+}={\frak{C}}_{\infty }^{+}W_{k,\bar{\mu}}(\rho ),\ \Delta
_{\infty }^{-}={\frak{C}}_{\infty }^{-}W_{-k,\bar{\mu}}(-\rho )
\end{equation}
where $\frak{C}_{\infty }^{+}$ and $\frak{C}_{\infty }^{-}$ are
normalization constants. The positive frequency mode at $\left| \rho \right|
\rightarrow \infty $ can be expressed in terms of the positive \ $\Delta
_{0}^{+}$ and negative $\Delta _{0}^{-}$ frequency modes via the Bogoliubov
transformation 
\begin{equation}
\Delta _{\infty }^{+}=\alpha \Delta _{0}^{+}+\beta \Delta _{0}^{-}.
\end{equation}
Since the Whittaker function $W_{k,\mu }(z)$ can be expressed in terms of $%
M_{k,\mu }(z)$ as follows: 
\begin{equation}
W_{k,\mu }(z)=\frac{\Gamma (-2\mu )}{\Gamma (\frac{1}{2}-\mu -k)}M_{k,\mu
}(z)+\frac{\Gamma (2\mu )}{\Gamma (\frac{1}{2}+\mu -k)}M_{k,-\mu }(z),
\end{equation}
we obtain that $\alpha $ and $\beta $ are given by the expressions 
\begin{equation}
\alpha =\frac{\frak{C}_{\infty }^{+}}{\frak{C}_{0}^{+}}\frac{\Gamma (-2\bar{%
\mu})}{\Gamma (\frac{1}{2}-\bar{\mu}-k)},\ \beta =-i\frac{\frak{C}_{\infty
}^{+}}{\frak{C}_{0}^{+}}\frac{\Gamma (2\bar{\mu})}{\Gamma (\frac{1}{2}+\bar{%
\mu}-k)}\exp (-\pi \left| \bar{\mu}\right| )  \label{alpha}
\end{equation}
then 
\begin{equation}
\left| \frac{\beta }{\alpha }\right| ^{2}=\left| \frac{\Gamma (\frac{1}{2}-%
\bar{\mu}-k)}{\Gamma (\frac{1}{2}+\bar{\mu}-k)}\right| ^{2}\exp (-2\pi
\left| \bar{\mu}\right| )  \label{rela}
\end{equation}
The computation of the density of particles created is straightforward from
Eq. (\ref{rela}) and the normalization condition of the wave function which,
for scalar particles, means that the Bogoliubov coefficients satisfy the
relation 
\begin{equation}
\left| \alpha \right| ^{2}-\left| \beta \right| ^{2}=1.  \label{norma1}
\end{equation}
Taking into account that \cite{Lebedev} 
\begin{equation}
\left| \Gamma (\frac{1}{2}+iy)\right| ^{2}=\frac{\pi }{\cosh \pi y},
\end{equation}
we find that the density of particles created is 
\begin{equation}
n=\left| \beta \right| ^{2}=\frac{\exp (\pi (\left| \bar{\mu}\right| +\left|
k\right| ))\cosh (\pi (\left| \bar{\mu}\right| +\left| k\right| ))\exp
(-2\pi \left| \bar{\mu}\right| )}{\sinh (2\pi \left| \bar{\mu}\right| )}
\label{dens1}.
\end{equation}
It is worth mentioning that, thanks to the relations (\ref{norma1}) and (\ref
{rela}), we did not have to compute the normalization constants $\frak{C}%
_{\infty }^{+}$, $\frak{C}_{\infty }^{-}$ and $\frak{C}_{0}^{+}$. \ \ When
the electric field is switched off, \ the parameter $k$ is zero and the
density of particles created (\ref{dens1}) reduces to a thermal distribution 
\cite{Bukhbinder}: 
\begin{equation}
n=\frac{1}{\exp 2\pi \sqrt{k_{\bot }^{2}+2\xi -\frac{1}{4}}-1}  \label{nesc}.
\end{equation}

\subsection{Dirac particles}

In order to compute the density of spin $1/2$ particles created, we proceed
to solve the Dirac equation in the cosmological background (\ref{1}) in the
presence of the electric field (\ref{A}).

The covariant Dirac equation in curved space in the presence of
electromagnetic fields can be written as follows 
\begin{equation}
\{\gamma ^{\mu }(\partial _{\mu }-\Gamma _{\mu }-ieA_{\mu })+m\}\Psi =0
\label{Dir}
\end{equation}
where the curved gamma $\gamma ^{\mu }$ matrices satisfy the anticommutation
relation $\left\{ \gamma ^{\mu },\gamma ^{\nu }\right\} _{+}=2g^{\mu \nu }$
and the spinor connections $\Gamma _{\mu }$ are 
\begin{equation}
\Gamma _{\mu }=\frac{1}{4}g_{\lambda \alpha }\left[ \left( \frac{\partial
b_{\nu }^{\beta }}{\partial x^{\mu }}\right) a_{\beta }^{\alpha }-\Gamma
_{\nu \mu }^{\alpha }\right] s^{\lambda \nu }  \label{spinc}
\end{equation}
where 
\begin{equation}
s^{\lambda \nu }=\frac{1}{2}(\gamma ^{\lambda }\gamma ^{\nu }-\gamma ^{\nu
}\gamma ^{\lambda }).
\end{equation}
The matrices $b_{\alpha }^{\beta }$, $a_{\beta }^{\alpha }$ establish the
connection between the curved $\gamma ^{\mu }$ and Minkowski $\tilde{\gamma}%
^{\mu }$ Dirac matrices as 
\begin{equation}
\gamma _{\mu }=b_{\mu }^{\alpha }\tilde{\gamma}_{\alpha },\mbox{ }\gamma
^{\mu }=a_{\beta }^{\mu }\tilde{\gamma}^{\beta }
\end{equation}
and  
\begin{equation}
\tilde{\gamma}^{\lambda }\tilde{\gamma}^{\nu }+\tilde{\gamma}^{\nu }\tilde{%
\gamma}^{\lambda }=2\eta ^{\lambda \nu }.
\end{equation}
Since the line element (\ref{1}) is diagonal, we choose to work in the
diagonal tetrad 
\begin{equation}
\gamma ^{\mu }=\sqrt{\left| g^{\mu \mu }\right| }\tilde{\gamma}^{\mu },\ \ %
\mbox{no sum}.  \label{conec}
\end{equation}
Substituting Eq. (\ref{conec}) into Eq. (\ref{spinc}) we obtain 
\begin{equation}
\Gamma _{1}=\frac{1}{2}\tilde{\gamma}^{0}\tilde{\gamma}^{1},\ \Gamma _{2}=%
\frac{1}{2}\tilde{\gamma}^{0}\tilde{\gamma}^{2},\ \Gamma _{3}=0,\ \Gamma
_{0}=0,  \label{con}
\end{equation}
and the Dirac equation (\ref{Dir}) takes the form 
\begin{equation}
\left\{ \tilde{\gamma}^{0}\frac{\partial }{\partial t}+\frac{1}{t}(\tilde{%
\gamma}^{1}\frac{\partial }{\partial x}+\tilde{\gamma}^{2}\frac{\partial }{%
\partial y})+\tilde{\gamma}^{3}(\frac{\partial }{\partial z}-\frac{ieC}{t}%
)+m\right\} \Psi _{0}=0  \label{Dira}
\end{equation}
where $\Psi _{0}=t\Psi .$ The factor $t$ was introduced in order to cancel
the contribution due to the spinor connections \ (\ref{con}). Equation (%
\ref{Dira}) can be written as a sum of two first order commuting
differential operators as follows \cite{Villalba3,Villalba4}: 
\begin{equation}
(\hat{K}_{1}+\hat{K}_{2})\Phi =0,
\end{equation}
\begin{equation}
\hat{K}_{2}\Phi =\emph{k}\Phi =-\hat{K}_{1}\Phi ,
\end{equation}
where the spinor $\Phi $ is related to $\Psi _{0}$ via the expression 
\begin{equation}
\tilde{\gamma}^{3}\tilde{\gamma}^{0}\Psi _{0}=\Phi ,\ 
\end{equation}
and $\emph{k}$ is a separation constant. The operators $\hat{K}_{1}$ and $\ 
\hat{K}_{2}$ read 
\begin{equation}
\hat{K}_{1}=t\left[ \gamma ^{3}\frac{\partial }{\partial t}+\gamma ^{0}(%
\frac{\partial }{\partial z}-i\frac{eC}{t})+m\gamma ^{3}\gamma ^{0}\right], 
\label{k1}
\end{equation}
\begin{equation}
\hat{K}_{2}=\left( \tilde{\gamma}^{1}\frac{\partial }{\partial x}+\tilde{%
\gamma}^{2}\frac{\partial }{\partial y}\right) \tilde{\gamma}^{3}\tilde{%
\gamma}^{0}.  \label{K2}
\end{equation}
Since Eq. (\ref{Dira}) commutes with $-i\nabla $, the spinor $\Phi $ can be
written $\ $as $\Phi =\Phi _{0}\exp (i(k_{x}x+k_{y}y+k_{z}z))$. Choosing to
work in the following representation of the Dirac matrices: 
\begin{equation}
\tilde{\gamma}^{0}=\left( 
\begin{array}{cc}
-i\sigma ^{1} & 0 \\ 
0 & i\sigma ^{1}
\end{array}
\right) ,\ \tilde{\gamma}^{1}=\left( 
\begin{array}{cc}
0 & i \\ 
-i & 0
\end{array}
\right) ,\ \tilde{\gamma}^{2}=\left( 
\begin{array}{cc}
\sigma ^{2} & 0 \\ 
0 & -\sigma ^{2}
\end{array}
\right) ,\ \tilde{\gamma}^{3}=\left( 
\begin{array}{cc}
\sigma ^{3} & 0 \\ 
0 & -\sigma ^{3}
\end{array}
\right),   \label{rep}
\end{equation}
we see that the equation $\hat{K}_{2}\Phi =\emph{k}\Phi $ helps us to
determine the relation between the components of the bispinor $\Phi $%
\begin{equation}
\Phi _{0}=\left( 
\begin{array}{c}
\Phi _{1} \\ 
\Phi _{2}
\end{array}
\right) =\left( 
\begin{array}{c}
\Phi _{1} \\ 
\frac{k_{x}\sigma ^{2}}{ik_{y}-\emph{k}}\Phi _{1}
\end{array}
\right)   \label{bisp}
\end{equation}
where 
\begin{equation}
\emph{k}=i\sqrt{k_{x}^{2}+k_{y}^{2}}.
\end{equation}
Using the representation (\ref{rep}) we find that Eq. (\ref{k1}) reduces
to the system of equations 
\begin{equation}
\left\{ \sigma ^{3}\frac{d}{dt}+\sigma ^{1}(k_{z}-\frac{eC}{t})+m\sigma ^{2}+%
\frac{\emph{k}}{t}\right\} \Phi _{1}=0,  \label{uno}
\end{equation}
\begin{equation}
\left\{ -\sigma ^{3}\frac{d}{dt}+-\sigma ^{1}(k_{z}-\frac{eC}{t})+m\sigma
^{2}+\frac{\emph{k}}{t}\right\} \Phi _{2}=0.  \label{dos}
\end{equation}
Taking into account the structure of the bispinor (\ref{bisp}), it is not
difficult to see that Eq. (\ref{dos}) is equivalent to Eq. (\ref{uno}) and
consequently we have reduced the problem of solving \ Eq. (\ref{Dira}) to
that of finding the solution of Eq. (\ref{uno}).

In order to solve Eq. (\ref{uno}) we proceed as follows. Since the structure
of Eq. (\ref{uno}) resembles the one obtained after solving the hydrogen
atom in spherical coordinates \cite{Akhiezer,Villalba5} and we are
interested in identifying positive and negative frequency modes, it would be
desirable to have solutions of Eq. (\ref{uno}), as in the scalar case, in
terms of single Whittaker functions. For this purpose, we apply to the
spinor $\Phi _{1}$ the linear transformation $T$ 
\begin{equation}
T\Phi _{1}=F=\left( 
\begin{array}{c}
f_{1} \\ 
f_{2}
\end{array}
\right)   \label{efe}
\end{equation}
with 
\begin{equation}
T=\left( 
\begin{array}{cc}
\sqrt{im+k_{z}} & -\sqrt{im-k_{z}} \\ 
\sqrt{im+k_{z}} & \sqrt{-im-k_{z}}
\end{array}
\right).   \label{T}
\end{equation}
The components of the spinor $F$ \ (\ref{efe}) satisfy the system of
equations 
\begin{equation}
\rho \frac{df_{1}}{d\rho }-f_{1}\left( \frac{-Cek_{z}}{\sqrt{-k_{z}^{2}-m^{2}%
}}+\frac{\rho }{2}\right) -f_{2}\left( \frac{IeCm}{\sqrt{-k_{z}^{2}-m^{2}}}%
-k\right) =0,  \label{eins}
\end{equation}
\begin{equation}
\rho \frac{df_{2}}{d\rho }+f_{2}\left( \frac{-Cek_{z}}{\sqrt{-k_{z}^{2}-m^{2}%
}}+\frac{\rho }{2}\right) +f_{1}\left( \frac{IeCm}{\sqrt{-k_{z}^{2}-m^{2}}}%
+k\right) =0,  \label{zwei}
\end{equation}
whose solutions can be expressed in terms of the Whittaker functions $W_{k,\mu
}(z)$ and $M_{k,\mu }(z)$ \cite{Lebedev,Abramowitz}.

In order to construct the positive and negative energy frequency modes, we
use the asymptotic behavior of the hypergeometric functions (\ref{M}) and (%
\ref{W}) and compare the solutions of Eqs. (\ref{eins}) and (\ref{zwei}) with
\ Eqs. (\ref{zero}) and (\ref{infi}), obtained after solving the Hamilton-Jacobi
equation. Using this procedure, we identify the positive frequency mode  as $%
t\rightarrow +\infty $ as

\ 
\begin{equation}
F_{+\infty }^{+}=\left( 
\begin{array}{c}
{\frak{C}_{\infty }}^{+} \rho ^{-1/2}W_{-\frac{1}{2}+\frac{eCk_{z}}{\sqrt{%
-k_{z}^{2}-m^{2}}},\sqrt{k^{2}-e^{2}C^{2}}}(\rho ) \\ 
\frac{1}{k-\frac{ICem}{\sqrt{-k_{z}^{2}-m^{2}}}}{\frak{C}}_{\infty }^{+} \rho
^{-1/2}W_{\frac{1}{2}+\frac{eCk_{z}}{\sqrt{-k_{z}^{2}-m^{2}}},\sqrt{%
k^{2}-e^{2}C^{2}}}(\rho )
\end{array}
\right) 
\end{equation}
with 
\begin{equation}
\mu =\sqrt{\emph{k}^{2}-e^{2}C^{2}}=i\sqrt{k_{\perp }^{2}+e^{2}C^{2}}=i%
\tilde{\mu}  \label{par}
\end{equation}
$\frak{C}_{\infty }^{+}$ is a normalization constant according to the
inner product 
\begin{equation}
\left\langle \Phi _{1},\Phi _{2}\right\rangle =\int_{\sigma }\bar{\Phi}%
_{1}^{{}}\gamma ^{\mu }\Phi _{2}\sqrt{-g}d\sigma _{\mu }  \label{normd}
\end{equation}
where $\sigma $ is an arbitrary spacelike hypersurface. Analogously, we
have that as $t\rightarrow 0$ the positive mode is 
\begin{equation}
F_{0}^{+}=\left( 
\begin{array}{c}
{\frak{C}_{0}}^{+}\rho ^{-1/2} M_{-\frac{1}{2}+\frac{eCk_{z}}{\sqrt{%
-k_{z}^{2}-m^{2}}},\mu }(\rho ) \\ 
-\frac{\frac{eCk_{z}}{\sqrt{-k_{z}^{2}-m^{2}}}+\mu }{\emph{k}-\frac{ICem}{%
\sqrt{-k_{z}^{2}-m^{2}}}}{\frak{C}_{0}}^{+}\rho ^{-1/2} M_{\frac{1}{2}+\frac{%
eCk_{z}}{\sqrt{-k_{z}^{2}-m^{2}}},\mu }(\rho )
\end{array}
\right) 
\end{equation}
where $\frak{C}_{0}^{+}$ is also a normalization constant. \ Using the
asymptotic behavior of the Whittaker function $M_{k,\mu }(z)$ (\ref{M}) and
the normalization condition (\ref{normd}), we obtain the result for the negative
frequency mode $F_{0}^{-}$ as $t\rightarrow 0$: 
\begin{equation}
F_{0}^{-}=\left( 
\begin{array}{c}
e^{\pi \tilde{\mu}}{\frak{C}}_{0}^{+}\rho ^{-1/2} M_{-\frac{1}{2}+\frac{eCk_{z}%
}{\sqrt{-k_{z}^{2}-m^{2}}},-\mu }(\rho ) \\ 
-\frac{\frac{eCk_{z}}{\sqrt{-k_{z}^{2}-m^{2}}}-\mu }{\emph{k}-\frac{ICem}{%
\sqrt{-k_{z}^{2}-m^{2}}}}e^{\pi \tilde{\mu}}{\frak{C}}_{0}^{+}\rho ^{-1/2} M_{%
\frac{1}{2}+\frac{eCk_{z}}{\sqrt{-k_{z}^{2}-m^{2}}},-\mu }(\rho )
\end{array}
\right) 
\end{equation}
The density of particles created can be calculated with the help of the
Bogoliubov transformation 
\begin{equation}
F_{\infty }^{+}=\alpha F_{0}^{+}+\beta F_{0}^{-},
\end{equation}
\begin{equation}
F_{+\infty }^{+}=\frac{\Gamma (-2i\tilde{\mu})}{\Gamma (\frac{1}{2}-i\tilde{%
\mu}-\frak{k})}F_{0}^{+}+\frac{\Gamma (2i\tilde{\mu})}{\Gamma (\frac{1}{2}+i%
\tilde{\mu}-\frak{k})}e^{-\pi \tilde{\mu}}F_{0}^{-},
\end{equation}
where ${\frak{k}} =-\frac{1}{2}+\frac{Ck_{z}}{\sqrt{-k_{z}^{2}-m^{2}}}.$ Using
the property of the Gamma function \cite{Lebedev} 
\begin{equation}
\left| \Gamma (iy)\right| ^{2}=\frac{\pi }{y\sinh \pi y}
\end{equation}
we obtain 
\begin{equation}
\left| \frac{\beta }{\alpha }\right| ^{2}=e^{-2\pi \tilde{\mu}}\frac{\sinh
\pi (\tilde{\mu}+\frac{eCk_{z}}{\sqrt{k_{z}^{2}+m^{2}}})}{\sinh \pi (\tilde{%
\mu}-\frac{eCk_{z}}{\sqrt{k_{z}^{2}+m^{2}}})}\frac{\left( \tilde{\mu}-\frac{%
eCk_{z}}{\sqrt{k_{z}^{2}+m^{2}}}\right) }{\left( \tilde{\mu}+\frac{eCk_{z}}{%
\sqrt{k_{z}^{2}+m^{2}}}\right) }.  \label{coc}
\end{equation}
From Eq. (\ref{coc}) and taking into account the normalization condition for
Dirac particles 
\begin{equation}
\left| \alpha \right| ^{2}+\left| \beta \right| ^{2}=1
\end{equation}
we find that the density of Dirac particles created is
\begin{equation}
\left| \beta \right| ^{2}=\frac{\exp (-2\pi \bar{\mu})\left( \tilde{\mu}-%
\frac{eCk_{z}}{\sqrt{k_{z}^{2}+m^{2}}}\right) \sinh \pi (\tilde{\mu}+\frac{%
eCk_{z}}{\sqrt{k_{z}^{2}+m^{2}}})}{\exp (-2\pi \tilde{\mu})\left( \tilde{\mu}%
-\frac{eCk_{z}}{\sqrt{k_{z}^{2}+m^{2}}}\right) \sinh \pi (\tilde{\mu}+\frac{%
eCk_{z}}{\sqrt{k_{z}^{2}+m^{2}}})+\left( \tilde{\mu}+\frac{eCk_{z}}{\sqrt{%
k_{z}^{2}+m^{2}}}\right) \sinh \pi (\tilde{\mu}-\frac{eCk_{z}}{\sqrt{%
k_{z}^{2}+m^{2}}})}  \label{den}
\end{equation}
In the absence of an electric field, the density of particles created (\ref{den}%
) becomes thermal: 
\begin{equation}
n=\frac{1}{\exp 2\pi \left| k_{\perp }^{{}}\right| +1}.
\end{equation}

\section{Discussion of the results}

From the expressions (\ref{dens1}) and (\ref{den}) we observe that, 
in the presence of the electric field (\ref{A}), the distribution of particles 
created is not
thermal and it strongly depends on the strength $eC.$ For very strong
electric fields we have that Eq. (\ref{dens1})  reduces to 
\begin{equation}
n\thicksim \exp (-2\pi \sqrt{k_{\bot }^{2}+e^{2}C^{2}+2\xi -\frac{1}{4}}+%
\frac{2\pi eCk_{z}}{\sqrt{k_{z}^{2}+m^{2}}}),  \label{dist}
\end{equation}
an analogous result can be obtained for spin $1/2$ particles. Expression (\ref
{dist}) \ corresponds to a thermal distribution with an effective mass and a
chemical potential proportional to $eC$ . \ Equation (\ref{dist}) \ shows that
the presence of strong electric fields  contributes significantly  to the
creation of particles. The results obtained in this article show that the
quasiclassical approach permits one to compute positive and negative
frequency modes even when spacetime and electromagnetic sources are not
static, and  encourage us to study quantum effects in more realistic
anisotropic scenarios.

\acknowledgments

One of the authors (V.M.V) acknowledges a fellowship from the Alexander von
Humboldt Stiftung.

\end{document}